\documentclass[twocolumn,prl,aps,showpacs]{revtex4}
\usepackage{amssymb}

\usepackage{amsmath}
\usepackage{graphicx}
\usepackage{dcolumn}
\usepackage{bm}

\begin{document}

\preprint{APS/123-QED}
\title{Theory of control of spin/photon interface for quantum networks}
\author{Wang Yao}
\author{Ren-Bao Liu}
\author{L. J. Sham}
\affiliation{Department of Physics, University of California San
Diego, La Jolla, California 92093-0319}
\date{\today}
\begin{abstract}
A cavity coupling a charged nanodot and a fiber can act as a
quantum interface, through which a stationary spin qubit and a
flying photon qubit can be inter-converted via cavity-assisted
Raman process. This Raman process can be controlled to generate or
annihilate an arbitrarily shaped single-photon wavepacket by
pulse-shaping the controlling laser field. This quantum interface
forms the basis for many essential functions of a quantum network,
including sending, receiving, transferring, swapping, and
entangling qubits at distributed quantum nodes as well as a
deterministic source and an efficient detector of a single photon
wavepacket with arbitrarily specified shape and average photon
number. Numerical study of noise effects on the operations shows
high fidelity.
\end{abstract}

 \maketitle

Quantum networks composed of local nodes and quantum channels are
essential for
quantum communication and desirable for scalable and distributed
quantum computation
\cite{Cirac_DistributedQC,DiVincenzo_Criteria_7}. Spins in quantum
dots \cite{Loss_QDspinQC} or stable levels of atoms or ions
\cite{Monroe_atom,Cirac_Ion} are good candidates for stationary
qubits which can be locally stored and manipulated
\cite{Imamoglu_CQED_Spin,ORKKY,Chen_Raman}. Photons in optical
fibers or waveguides are ideal flying qubits for carrying quantum
information between the local nodes. A quantum interface
inter-converting local and flying qubits is the key component of
the quantum network. A very recent experiment demonstrating
entanglement between photon polarizations and states in an atom
\cite{monroe} represents a breakthrough in this direction. Another
milestone toward quantum networks is the proposal of Cirac {\it et
al.} based on cavity-assisted Raman processes
\cite{StateTransfer_Cirac_Kimble}, which employs time-symmetrical
carrier pulses and mutually time-reversed operations at two nodes
to transfer a qubit state from one node to another.

This Letter proposes a faithful and controllable spin/photon
quantum interface without imposing the symmetry requirement
between two nodes.  It also describes a semiconductor quantum dot
and cavity operation for this quantum interface and, from it, the
architecture of a solid-state quantum network.

We note that the state-transfer process proposed in
Ref.~\onlinecite{StateTransfer_Cirac_Kimble} can be separated into
two steps: the sending operation at one node which maps a
stationary qubit into a flying qubit by the evolution
$[C_g|g\rangle+C_e|e\rangle]\otimes |{\rm vac}\rangle\rightarrow
|g\rangle\otimes[C_g|{\rm vac}\rangle+C_e|\alpha(t)\rangle]$, and
the receiving operation at another node which maps the flying
qubit into a stationary one, where $|g\rangle$ and $|e\rangle$ are
the stationary qubit states and the flying qubit is represented by
the vacuum state $\left|{\rm vac}\right\rangle$ and a
single-photon state with wavepacket $\alpha(t)$. We will show that
both the sending and receiving processes can be independently
controlled by shaping the laser pulses. Two aspects of the process
are controllable: the production of an arbitrarily shaped pulse
provided that it is sufficiently smooth and the operation of the
Raman process as a partial cycle, in which the initial state
$|e\rangle\otimes|{\rm vac}\rangle$ is mapped into an entangled
state $\cos\theta|e\rangle\otimes|{\rm
vac}\rangle+\sin\theta|g\rangle\otimes|\alpha(t)\rangle$ for any
$\theta\in [0,\pi/2]$.

With such controllability in hand, this quantum interface can
accomplish many essential functions of a quantum network: (i) It
can send a flying quantum state and can also function as a
deterministic source of single-photons with arbitrary pulse shape
and controllable photon number. (ii) It can receive a flying
quantum state, being an efficient single-photon detector provided
that the incoming photon pulse shape is known. (iii) The sending
and receiving processes combined transfer a state from one node to
another. (iv) An incoming flying qubit may be swapped with a
stationary qubit which enables the swap of two remote qubits. (v)
An entangled state of the stationary and flying qubits is produced
in a partial Raman cycle. (vi) Two stationary qubits separated far
away are entangled when the photon state generated by the partial
Raman cycle is mapped into a stationary qubit.

\begin{figure}[t]
\includegraphics[width=7.8cm, height=7.5cm, bb=55 165 575 665,
clip=true]{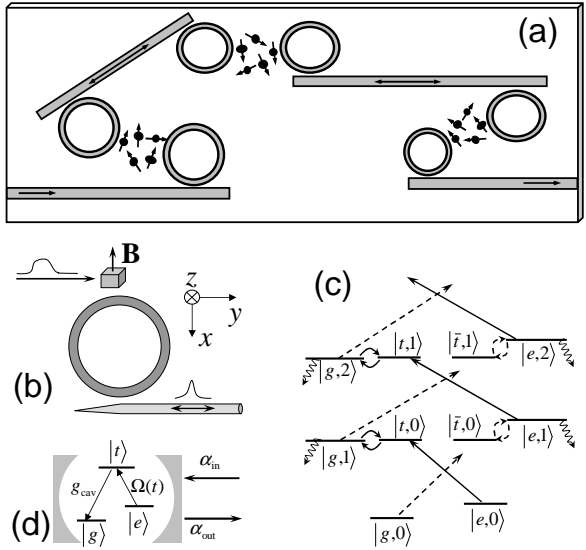} \caption{(a) The schematics of the solid-state
quantum network constructed by doped quantum dots, micro-rings,
and waveguides.  (b) A high-$Q$ micro-ring coupling a `tapered'
waveguide and a doped quantum dot. (c) The level diagram and
optical process. Straight, curved, and wavy arrows represent the
laser excitation, dot-cavity coupling, and cavity-fiber tunneling,
respectively.  The resonant and off-resonant processes are
represented by solid and dashed lines, respectively. (d) The
simplified cavity-assisted Raman process.} \label{system1}
\end{figure}

A substantial list of recent experimental advances on optical
manipulation of excitons in single quantum dots
\cite{Steel_QDgate}, nanodot-microsphere coupling
\cite{WangHL_MCdot}, cavity-fiber coupling \cite{vahala_review},
and fabrication of high-quality microcavities and waveguides, both
on semiconductor surfaces \cite{vahala_review} and in photonic
crystals \cite{2D_photoniclattice,Akahane_PhotonicCrystal}, makes
it fairly realistic for us to outline a physical plan for a
solid-state quantum network for scalable and distributed
processing of quantum information [see Fig.~\ref{system1} (a)]. In
this network structure, a local node is formed by a cluster of
charged quantum dots, and the electron spins representing
stationary qubits, having memory lifetime $\sim 10$ $\mu$s
\cite{DaSSarmaSpin2}, can be optically manipulated in the time
scale $\sim 10$ ps \cite{Imamoglu_CQED_Spin,ORKKY,Chen_Raman}.
Fibers or waveguides can act as sufficiently lossless quantum
channels for flying photons. Since it takes less than 1 ns for a
photon to travel 10 cm, the quantum network can be easily scaled
from nano-devices to centimeter-sized chips. The loss or
decoherence of photons in fibers is negligible in such a short
distance and may be ameliorated by quantum repeaters in the long
distance transfer \cite{duan}. The quantum interface is made up of
a high-$Q$ microcavity coupling a quantum dot and a  fiber or
waveguide [see Figure~\ref{system1}(b)]. The dot-cavity-fiber
structure also allows the ultrafast initialization and
non-destructive readout of the spin qubit in the dot
\cite{ReadWrite}, essential for quantum error correction and
scalable quantum computation.

The detailed optical process involved in the quantum interface is
depicted in Fig.~\ref{system1}(c). The qubit is represented by the
two spin states $|g\rangle$ and $|e\rangle$ which is split by a
static magnetic field.  The intermediate states in the Raman
process are the two degenerate trion states $|t\rangle$ and
$|\bar{t}\rangle$. The geometrical configuration and optical
polarizations \cite{SolidstatephaseGate} are set such that the
cavity mode of frequency $\omega_c$ couples only the transitions
$|g\rangle \rightarrow |t\rangle$ and $|e\rangle \rightarrow
|\bar{t}\rangle$ with coupling constant $g_{\rm cav}$, and the
controlling laser of frequency $\omega_L$ couples only the
transitions $|e\rangle \rightarrow |t\rangle$ an $|g\rangle
\rightarrow |\bar{t}\rangle$ with the complex Rabi frequency
$\Omega(t)$. The Raman process satisfies the resonance condition:
$\omega_{t}=\omega_L+\omega_{e}=\omega_{c}+\omega_{g}$. By the
Zeeman splitting and the selection rules, the trion state
$|\bar{t}\rangle$ is off-resonantly coupled to the laser beam and
the cavity modes (shown by dashed lines in the figure). The
cavity mode is coupled to the fiber continuum  by the coupling
constant $\kappa$.

The basic Raman process is quite simple. At a sending node, the
laser field first resonantly excites the spin state $|e,0\rangle$
to the trion state $|t,0\rangle$, then the trion state is
resonantly coupled to the cavity state $|g,1\rangle$, and finally
the cavity state jumps to the spin state $|g,0\rangle$ by emitting
a photon into the quantum channel. The receiving mode is just the
time reversed process. The off-resonant excitation which leads to
rather complicated dynamics involving multi-photon states  in the
cavity is rendered negligible by a strong magnetic field making
the Zeeman splitting much larger than the cavity-dot coupling and
the Rabi frequency.  As shown in Fig.~\ref{system1}(d), the
optical process is then simplified to a cavity-assisted resonant
Raman process in a $\Lambda$-type three-level system. In this
simplified system, we can find,  for any designated shape and
average photon number of the emitted single-photon wavepacket, an
analytical solution of the pulse shape of the laser field (or Rabi
frequency). With this analytically obtained Rabi frequency as the
controlling input, numerical calculations including the
non-resonant transitions and realistic decoherence have been
performed and high fidelity of desired operations at the quantum
interface is demonstrated.

We first present the analytical solution of the laser pulse for
the controlled Raman transition converting a spin qubit state
$C_{g}\left| g,0\right\rangle\left| {\rm vac}\right\rangle +C_{e}
\left| e,0\right\rangle \left| {\rm vac}\right\rangle$ to a flying
photon state.  We note that the simplified system depicted in
Fig.~\ref{system1}(d), under the optical excitation and the
cavity-dot and cavity-channel interaction, has two invariant
Hilbert subspaces, with the basis $\left\{
\left|g,0\right\rangle\left|{\rm vac}\right\rangle\right\}$ and
$\left\{\left|e,0\right\rangle \left| {\rm vac}\right\rangle,
\left| t,0\right\rangle \left| {\rm vac}\right\rangle, \left|
g,1\right\rangle \left| {\rm vac}\right\rangle, \left|g,0
\right\rangle\left|\omega\right\rangle\right\}$, respectively
(where $\left|\omega\right\rangle$ denotes the one-photon Fock
state of the fiber mode with frequency $\omega$). So the evolution
of the system can be described by the state $C_{g}\left|
g,0\right\rangle\left| {\rm vac}\right\rangle +C_{e} \left|
\Psi^e\left(t\right)\right\rangle$, where
\begin{eqnarray}
\left| \Psi^e \left( t\right) \right\rangle &=& \beta _{e}\left(
t\right) \left| e,0\right\rangle \left| {\rm vac}\right\rangle  +
\beta _{t}\left( t\right) \left| t,0\right\rangle \left| {\rm
vac}\right\rangle \nonumber \\ &+& \beta_{c}\left( t\right) \left|
g,1\right\rangle \left| {\rm vac}\right\rangle +\int_{0}^{\infty
}d\omega\alpha_{\omega}\left( t\right) \left| g,0
\right\rangle\left|\omega\right\rangle. \nonumber
\end{eqnarray}
Within the Weisskopf-Wigner approximation, the equation of motion
for the resonant Raman process in the interaction picture can be
derived as
\begin{subequations}
\begin{eqnarray}
\dot{\beta}_{e} &=&- \Omega ^{\ast }(t) \beta_{t} /2 ,\label{eom1_1} \\
\dot{\beta}_{t} &=&+  \Omega(t) \beta_{e}/2+g^*_{\rm cav}\beta_{c} ,\label{eom1_2} \\
\dot{\beta}_{c} &=& - \gamma\beta_{c} /2-g_{\rm
cav}\beta_{t}-\sqrt{2\pi}\kappa^* \alpha _{\rm in}\left( t\right) \label{eom1_3} \\
 &=& + \gamma\beta_{c} /2-g_{\rm
cav}\beta_{t}-\sqrt{2\pi}\kappa^*\alpha _{\rm out}\left( t\right),
\label{eom1_3a}
\end{eqnarray}
\label{eom1}
\end{subequations}
where $\gamma \equiv 2 \pi \left|\kappa\right|^2 $ gives the
cavity damping rate, and $\alpha _{\rm in}\left( t\right)\equiv
\int d\omega \alpha_{\omega}\left( t_{0}\right)
e^{-i(\omega-\omega_c) t}/\sqrt{2\pi }$ with $t_0\rightarrow
-\infty$ and $\alpha_{\rm out}\left( t\right)\equiv \int
d\omega\alpha_{\omega}\left( t_1\right) e^{-i(\omega-\omega_c)
t}/\sqrt{2\pi }$ with $t_1\rightarrow +\infty$ are the incoming
and outgoing pulse of the photon in the quantum channel,
respectively. The quantum fluctuation caused by the quantum
channel is on the order of $\gamma/\omega_c\ll 1$ and thus the
Weisskopf-Wigner approximation is well justified here.

In the process of mapping the spin qubit to the photon qubit,
there is no incoming photon, so the initial conditions are:
$\alpha_{\rm in}(t)=0$, $\beta_c(t_0)=0$, $\beta_e(t_0)=1$ and
$\beta_t(t_0)=0$. Eq.~(\ref{eom1}) can be analytically solved as
\begin{subequations}
\begin{eqnarray}
\left|{\beta}_{e}\right| ^{2}&=&1-\sin^2\theta\int_{t_0 }^{t}
\left|
\tilde{\alpha}_{\rm out}\left( \tau \right) \right| ^{2}d\tau \nonumber \\
&& -\left| {\beta}_{c}\right| ^{2} - \left| g_{\rm
cav}\right|^{-2} \left| \dot{{\beta}}_{c}+{\gamma } {\beta}_{c} /2
\right| ^{2},
\label{bd1} \\
{d\over dt} \arg\left(\beta_e\right) &=&
{\left|\dot{{\beta}}_{c}+{\gamma }{\beta}_{c}/2\right|^2 \over
\left|g_{\rm cav}\right|^{2}\left|\beta_e\right|^{2}} {d\over
dt}\arg \left( \dot{{\beta}}_{c}+{\gamma
}{\beta}_{c}/2\right)\nonumber \\ && -
\left|\beta_e\right|^{-2}\left|\beta_c\right|^2 {d\over dt}\arg
\left(\beta_c\right)
 ,  \label{phase}\\
{\Omega\over 2} &=&-{g^*_{\rm cav}\beta_c\over
\beta_e}-{\ddot{\beta}_{c}+\gamma\dot{\beta}_c/2\over g_{\rm
cav}\beta_e}, \label{pulse}
\end{eqnarray}
\label{eom2}
\end{subequations}
where $\tilde{\alpha}_{\rm out}$ is the normalized wavepacket of
the emitted photon, and $\sin^2\theta$ is the average photon
number. For a photon number and a pulse shape arbitrarily
specified, the amplitude of the cavity mode is determined by
Eqs.~(\ref{eom1_3}) and (\ref{eom1_3a}) as $\beta_c=
\tilde{\alpha}_{\rm
out}\sin\theta/\left(\sqrt{2\pi}\kappa\right)$, and in turn the
controlling laser pulse $\Omega(t)$ can be analytical obtained
from Eq.~(\ref{pulse}). Note that the solution exists only when
the right-hand side of Eq.~(\ref{bd1}) is positive which requires
the specified output pulse be sufficiently smooth, i.e., the pulse
generation process be slower than the cavity decay and the
dot-cavity tunnelling (with time scales $\gamma^{-1}$ and $g_{\rm
cav}^{-1}$, respectively). At the remote future time
$t_1\rightarrow +\infty$, the photon emission process is
completed, i.e., $\beta_c(t_1)=\dot{\beta}_c(t_1)=0$, so
$\beta_e(t_1)=e^{i\phi}\cos\theta$ with the controllable phase
$\phi$ independent of the initial superposition (see
Eq.~(\ref{phase})). Compared to
Ref.~\onlinecite{StateTransfer_Cirac_Kimble}, the solution here is
not limited in the perturbative regime, and thus enables the
design of ultrafast operation.

When the full Raman transition is completed, $\theta=\pi/2$ and
$\beta_e(t_1)=0$, thus the spin qubit is mapped into the flying
qubit by the transformation
\begin{eqnarray}
\left(C_g|g\rangle+C_e|e\rangle\right) \otimes |{\rm
vac}\rangle\stackrel{\Omega}{\rightarrow}
|g\rangle\otimes\left[C_g|{\rm vac}\rangle+C_e|\tilde{\alpha}_{\rm
out}\rangle\right].
\end{eqnarray}
The mapping operation also functions as deterministic generation
of a single-photon wavepacket with any desired pulse shape
$\tilde{\alpha}_{\rm out}$ and average photon number
$\left|C_e\right|^2$.

The Raman cycle can also be controlled such that $\theta<\pi/2$.
In this case, the initial state $|e\rangle\otimes |{\rm
vac}\rangle$ is transformed into an entangled state of the
stationary spin and the flying photon
\begin{eqnarray}
e^{i\phi}\cos\theta|e\rangle\otimes|{\rm
vac}\rangle+\sin\theta|g\rangle\otimes|\tilde{\alpha}_{\rm
out}\rangle.
\end{eqnarray}
The entanglement entropy $-\cos^2\theta \log_2
\cos^2\theta-\sin^2\theta \log_2 \sin^2\theta$ can be set any
value between 0 and 1 depending on the rotating angle $\theta$.

The Raman process at the receiving node is basically the
time-reversal of the sending process. With the initial spin state
$|g\rangle$ and the incoming photon state $C_g|{\rm
vac}\rangle+C_e|\alpha_{\rm in}(t)\rangle$, the mapping
transformation is
\begin{eqnarray}
|g\rangle\otimes(C_g|{\rm vac}\rangle+C_e|\alpha_{\rm in}\rangle)
\stackrel{\Omega}{\rightarrow} (C_g|g\rangle+C_e|e\rangle)\otimes
|{\rm vac}\rangle \label{map2}.
\end{eqnarray}
As in the sending process, the incoming photon pulse $\alpha_{\rm
in}(t)$ can be arbitrarily specified provided that it is smooth
enough, and the photon is absorbed without reflection. As the spin
state converted from the photon state can be read out
non-destructively \cite{ReadWrite}, the receiving node can also
act as an efficient photon detector which measures the photon
number state given the photon pulse shape is known.

For swap operations, the waveguide connecting the two nodes should
be long enough to make the photon travelling time longer than the
operation time. If the operation at the sending node  has been
designed to produce an entangled state of the spin and the photon,
the mapping process at the receiving node will just produce a
non-locally entangled state of the two spins by the transformation
\begin{eqnarray}
&& |e\rangle_{1}|g\rangle_{2}
\otimes|{\rm vac}\rangle \notag  \\
&\stackrel{\Omega_1} {\rightarrow}
&e^{i\phi}\cos\theta|e\rangle_1|g\rangle_{2}\otimes|{\rm
vac}\rangle+\sin\theta|g\rangle_1|g\rangle_{2}\otimes|\tilde{\alpha}_{\rm
out}\rangle \nonumber \\
&\stackrel{\Omega_2}{\rightarrow} &
\left[e^{i\phi}\cos\theta|e\rangle_{1}|g\rangle_{2}+\sin\theta|g\rangle_1|e\rangle_{2}\right]
\otimes|{\rm vac}\rangle. \label{entangle}
\end{eqnarray}

The fidelity of all the quantum operations described above can be
evaluated by numerical simulations including the undesired
non-resonant dynamics, and unavoidable decoherence. The
decoherence results mainly from the trion decay due to spontaneous
emission and the intrinsic loss of the cavity mode, while the
fiber loss and the spin relaxation, though unavoidable, are
negligible on the time-scale ($\sim 100$ ps) and the space-scale
($\sim 1$ cm) considered here. The trion decay rate is chosen to
be a realistic value of $\Gamma=3 \mu$eV, and the intrinsic loss
rate of a high-$Q$ cavity is assumed to be $\gamma_0=0.1$ $\mu$eV
(corresponding to a $Q$-factor $\sim 10^7$). The cavity-fiber
tunnelling rate is set $\gamma=0.2$ meV and the dot-cavity
coupling constant is chosen as $g_{\rm cav}=0.1$ meV. The Zeeman
splitting of the spin under a quantizing magnetic field is 1 meV,
chosen much larger than the Rabi frequency and the cavity-dot
coupling in order to suppress the undesired effects: the
non-resonant excitation of the multi-photon states and AC Stark
shift of the energy levels. Numerical test shows that including up
to 3-photon states is enough to obtain convergent results. While
the multi-photon states of fiber modes can be projected into the
error subspace in the state-sending operation, they can still
carry quantum information in the state-transfer or non-local
entanglement operations, thus the quantum trajectory method
\cite{StateTransfer_Cirac_Kimble,CascadeOpenSystem_Carmichael} has
been employed in these cases to include the effect of multi-photon
pulses propagating in the fiber.

\begin{figure}[t]
\includegraphics[width=7.36cm, height=5.44cm, bb=45 235 505 575,
clip=true]{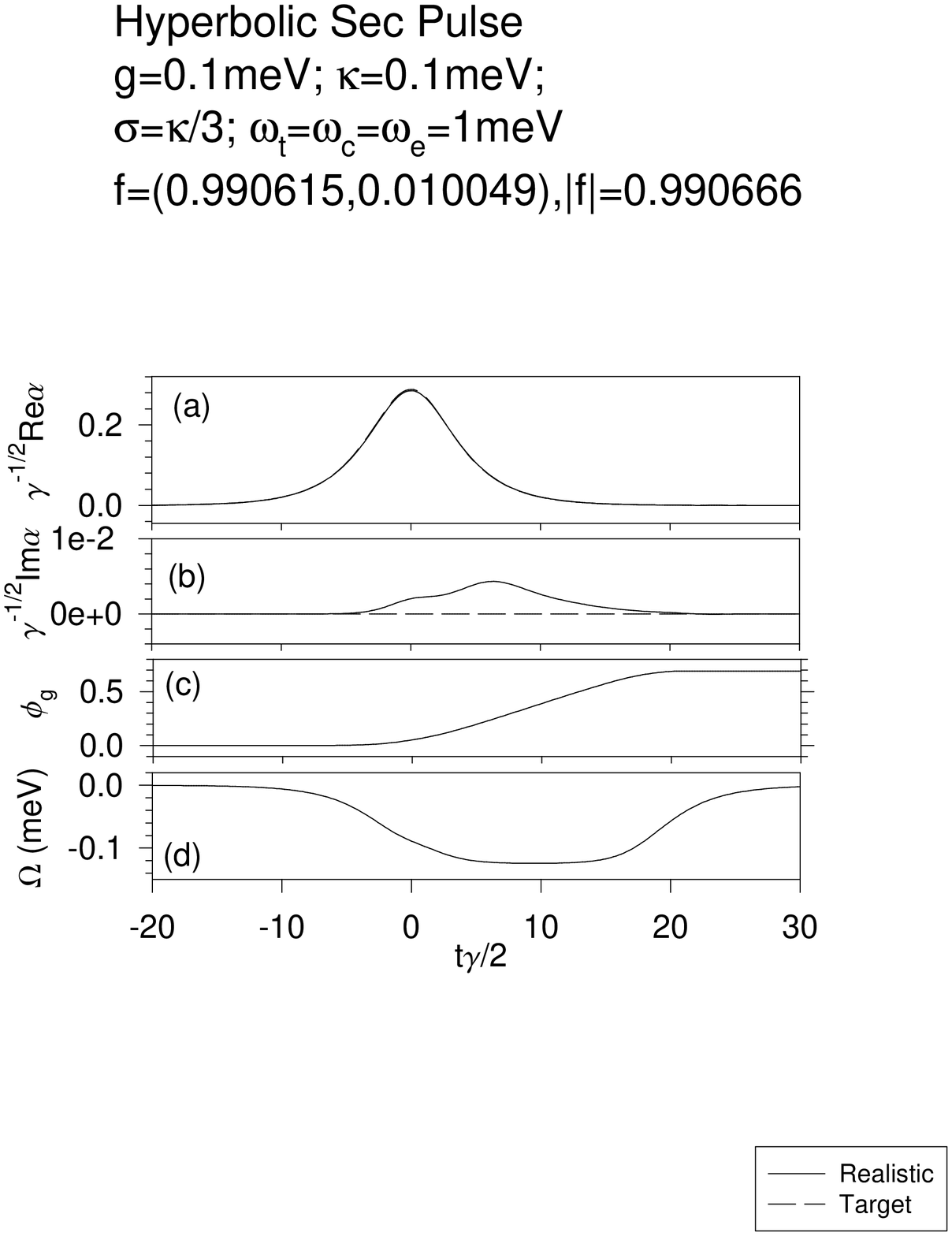} \caption{Illustration of the generation of a
sech-pulse. (a) Real part of the dimensionless amplitude of the
generated pulse (solid line) and target pulse (dashed line) as a
function of the dimensionless time $\gamma t/2$. (The difference
is not visible.) (b) Imaginary part of the generated pulse (solid
line) and target pulse (dashed line). (c) Phase drift of the state
$|g,0\rangle$. (d) Rabi frequency of the laser.} \label{gene}
\end{figure}

To demonstrate the efficiency of the quantum interface, we present
in Fig.~\ref{gene} the simulation result of mapping a spin state
to a flying photon wavepacket with the pulse shape designated to
be a sech-function as $\alpha^{\rm
ideal}_{\text{out}}(t)=\sqrt{\gamma/12}\,\text{sech}(\gamma t/6)$.
The actual final state of the evolution can be generally written
as $ C'_{g}e^{i\phi_g}\left| g\right\rangle\left| {\rm
vac}\right\rangle +C'_{e}\left|g\right\rangle \left|\alpha_{\rm
out} \right\rangle+|\varphi_{\text{err}}(t)\rangle.$  Due to the
non-resonant excitation, the wavefunction can have an error part
$|\varphi_{\text{err}}\rangle$ which lies outside the
qubit-subspace. The large Zeeman splitting and the smooth
switch-on and off of the laser field make this error less than
$0.12\%$. Another error due to the non-resonant coupling is the AC
Stark shift of the energy levels, which induces a phase drift  of
the ground state $|g\rangle$. This phase drift ($\phi_g$),
however, is independent of the coefficients $C_{g(e)}$ as the two
excitation pathways starting respectively from $|g\rangle$ and
$|e\rangle$ are independent of each other [see Fig.~\ref{system1}
(c)], so it can be compensated by a trivial single-qubit
operation. The fidelity of the pulse generation $|\langle
\alpha^{\rm ideal}_{\rm out}|\alpha_{\rm out}\rangle|\approx
0.9907$ is high. Including the leakage into the error subspace,
the overall fidelity of the mapping operation is still larger than
$0.99$. Due to the non-perturbative optical pumping and dot-cavity
coupling, the whole mapping process can be completed within 300
ps. The simulation of the photon absorption process shows an
overall fidelity greater than $0.99$ as well.

In the simulations of state-transfer and non-local entanglement,
the two remote nodes are assumed identical, but it should be noted
that in our scheme the two nodes can be controlled independently
and thus can be designed with different parameters if desired. The
state transfer process has been simulated with various carrier
pulse shapes. The overall fidelity, with the phase drift
compensated, is greater than $98\%$ for a sech-pulse
$\alpha(t)=\sqrt{\gamma/8} \text{sech}{(\gamma t/4)}$. Using the
same carrier pulse, we have also simulated the creation of
non-local entanglement. With the Rabi frequency analytically
designed for a Bell state $\left(e^{i\phi}|g\rangle _{1}|e\rangle
_{2}+ |e\rangle _{1}|g\rangle _{2}\right)/\sqrt{2}$, the actual
final state of the simulation turns out to be $0.6908
e^{i\phi}|g\rangle_{1}|e\rangle _{2}+ 0.7100|e\rangle
_{1}|g\rangle _{2}+0.1366|\varphi_{\text{err}}\rangle$. While the
probability of leakage out of the qubit subspace is less than
$1.87\%$, the spins are entangled with entanglement entropy
$0.9995$ and fidelity $0.9905$.

In summary, controllable and high-fidelity quantum interface for
inter-converting spin/photon qubits, based on cavity-assisted
Raman process, can be used to construct a scalable quantum network
with solid-state elements including charged quantum dots,
micro-rings (or microcavities), and waveguides (or fibers). The
theory assumes a knowledge of the parameters of the coupled
system. It paves the way for further exploration of quantum
feeback \cite{wiseman93} and learning algorithms \cite{rabitz00}
for this system.

This Work was supported by  NSF DMR-0099572, ARDA/ARO
DAAD19-02-1-0183, and QuIST/AFOSR F49620-01-1-0497.

\end{document}